\documentstyle[12pt]{article}
  \topmargin - 0.5 cm 
\begin{document}
  \begin{titlepage}
  \scrollmode

\begin{center}
{\Large\bf An Alternative Basis for the Wigner-Racah} 

\vspace{0.5cm}
{\Large\bf Algebra of the Group SU(2)}
\end{center}

\vspace{0.5cm}

\begin{center}
{\bf M.~KIBLER$^1$ and M. DAOUD$^2$}
\end{center}

\begin{center}
{$^1$Institut de Physique Nucl\'eaire de Lyon}\\
{IN2P3-CNRS et Universit\'e Claude Bernard}\\
{43 bd du 11 novembre 1918,}
{69622 Villeurbanne Cedex, France}
\end{center}

\begin{center}
{$^2$Laboratoire de Physique Th\'eorique}\\
{Universit\'e Mohammed V}\\
{Avenue Ibn Batouta, B.P. 1014, Rabat, Morocco}
\end{center}

\vspace{2cm}
\begin{abstract}

The Lie algebra of the classical group SU(2) is constructed from two quon 
algebras for which the deformation parameter is a common root of unity. This 
construction leads to (i) a (not very well-known) polar decomposition of the 
generators $J_-$ and $J_+$ of the SU(2) Lie algebra and to (ii) an alternative 
to the $\{ J^2, J_3 \}$ quantization scheme, viz., the $\{ J^2, U_r \}$ 
quantization scheme. The key ideas for developing the Wigner-Racah 
algebra of the group SU(2) in the $\{ J^2, U_r \}$ scheme are given. In 
particular, some properties of the coupling and recoupling coefficients as well 
as the Wigner-Eckart theorem in the $\{ J^2, U_r \}$ scheme are briefly 
discussed. 

\end{abstract}

\vskip 5.65 true cm
\noindent 
Paper to appear in the Proceedings of the International Conference 
{\bf Quantum Groups, Deformations and Contractions} 
(Istanbul, Turkey, 17 - 24 September 1997). 
The Proceedings of the Conference will be published in the 
Turkish Journal of Physics. 

\vfill
\thispagestyle{empty}
\end{titlepage}

\newpage

\section{Motivations and Introduction}
In recent years, intermediate statistics and deformed statistics were the
object of considerable interest [1-19]. The use of deformed oscillator algebras
proved to be useful in parastatistics, anyonic statistics and deformed
statistics. In particular, one- and two-parameter
deformations of the Bose-Einstein statistics (more precisely, deformations 
of the relevant second quantization formalism) were studied by several
authors [6-19]. A common characteristics of most of these studies is that it is
possible to obtain a Bose-Einstein condensation of a free gas of bosons 
in $D=2$ and 3 dimensions. However, in $D=3$ dimensions, the $q$-deformed
Bose-Einstein (B-E) temperature is generally greater than the classical
(corresponding to $q=1$) B-E temperature. In the specific case of 
$^4$He super-fluid in phase II, 
the usual $q$-deformations, i.e., the {\em \`a la} Biedenharn [20] and 
{\em \`a la} Macfarlane [21] 
$q$-deformations, yield the following inequality~: 
  \begin{eqnarray*}
  \left( T_{{\rm B-E}} \right)_{q \ne 1} >
  \left( T_{{\rm B-E}} \right)_{q  =  1} >
  \left( T_{{\rm B-E}} \right)_{{\rm exp}}  
  \end{eqnarray*}
so that we do not gain anything when passing from $q=1$ to $q \ne 1$. 
On the other hand, by using a {\em \`a la} Rideau [22,23] deformation, it is
feasible to lower the critical temperature $\left( T_{{\rm B-E}} \right)_{q \ne 1}$ 
due to the occurrence of a second parameter $\nu_0'$ in addition to the
deformation parameter $q$. This result corresponds to the model M$_1$
introduced in ref.[19]. For this model, we can obtain couples ($\nu_0',q$) for
which $\left( T_{{\rm B-E}} \right)_{q \ne 1}$ is in agreement with the experimental
value $\left( T_{{\rm B-E}} \right)_{{\rm exp}} \sim 2.17$~K. 
However, as a drawback, the model M$_1$ depends on two parameters. 
Although it is possible to find a physical
interpretation (in terms of the chemical potential) of the deformation
parameter $q$, there is up to now no satisfying interpretation of the
phenomenological parameter $\nu_0'$. 

The just mentioned difficulty to interpret the parameter $\nu_0'$ 
was the starting point of an investigation 
of alternative deformations of the second quantization formalism.  
More specifically, we investigated the {\em \`a la} Arik and Coon [24]
deformation but in the case where $q$ is a root of unity. (In the original 
work by Arik and Coon, the deformation parameter $q$ is a real number~: The
reality of $q$ ensures that the creation and annihilation operators 
are connected via Hermitean conjugation.) Thus,
we arrived at the conclusion that it is necessary to simultaneously consider
two quon algebras $A_q$ and $A_{\bar q}$ in order to obtain a convenient
framework for obtaining B-E condensation of quons. 

As a first by-product, we were naturally left to the definition and study of 
operators, referred to as $k$-fermion operators, that interpolate
between boson and  fermion  operators. These new operators arise through the 
consideration of two non-commuting quon algebras $A_q$ and $A_{\bar q}$ 
for which $q = {\rm exp}( { 2 \pi {\rm i} / k } )$ with 
$k \in {\bf N} \setminus \{ 0,1 \}$. 
The case $k=2$ corresponds to fermions and
the limiting case $k \to \infty$ to bosons. 
Generalized coherent states
(connected to $k$-fermionic states) and super-coherent states 
(involving a  $k$-fermionic sector and a purely bosonic sector) were 
examined. In addition, the operators in the $k$-fermionic algebra were 
used to find 
realizations of the Dirac quantum phase operator and of the $W_{\infty}$ 
Fairlie-Fletcher-Zachos  algebra [25]. All these matters were discussed in
Bregenz  (at the Symposium {\em Symmetries in Science X}), 
Dubna    (at the VIII International Conference on 
{\em Symmetry Methods in Physics}) and
Istanbul (at the International Workshop 
{\em Quantum Groups, Deformations and Contractions}) 
and shall be reported elsewhere [26,27]. 

In the present paper, we would like to deal with a second by-product of our
quon approach. Here, instead of considering two non-commuting quon algebras 
$A_q$ and $A_{\bar q}$, we shall consider two realizations of two 
commuting quon algebras 
corresponding to the same root of unity 
$q = {\rm exp}( { 2 \pi {\rm i} / k } )$ with 
$k \in {\bf N} \setminus \{ 0,1 \}$. We shall see 
how to construct (in Section 2) the Lie algebra of SU(2) from these two quon 
                                                                 algebras~; 
how to obtain    (in Section 3) an alternative to the $\{ J^2, J_z \}$ scheme 
                                                                 of SU(2)~; and 
how to develop   (in Section 4) the Wigner-Racah algebra of SU(2) in this new 
                                                                    scheme. 
In a  last  section (Section 5), we shall indicate some perspectives and briefly
discuss some open problems. 

\section{A Quon Approach to SU(2)}
We start with two commuting quon algebras $A_i = \{ a_{i-}, a_{i+}, N_i \}$, 
with $i = 1$ and $2$, for which the generators satisfy
  \begin{equation}
   a_{i-}a_{i+} - 
  qa_{i+}a_{i-} = 1, \quad \left[ N_i, a_{i\pm} \right] = \pm a_{i\pm}
  \end{equation}
where the deformation parameter 
  \begin{equation}
  q = \exp \left( {2 \pi {\rm i} \over k} \right) \quad {\rm with} \quad 
  k \in {\bf N} \setminus \{ 0,1 \} 
  \end{equation}
(the same for $A_1$ and $A_2$) is a root of unity. As constraint relations, 
compatible with (1) and (2), we take the nilpotency conditions 
  \begin{equation}
  \left( a_{i+} \right)^k = \left( a_{i-} \right)^k = 0 \quad 
  {\rm with} \quad k \in {\bf N} \setminus \{ 0,1 \} 
  \end{equation}
Grassmannian realizations of eqs.(1) and (3) are obtainable from ref.[26]. 
In this work, we take the representations of $A_1$ and $A_2$ defined by
  \begin{eqnarray*}
  a_{1+} |n_1) = |n_1 + 1), \quad 
  a_{1+} |k-1) = 0 
  \end{eqnarray*}
  \begin{eqnarray*}
  a_{1-} |n_1) = \left[ n_1   \right]_q |n_1-1), \quad  
  a_{1-} |0)   = 0
  \end{eqnarray*}
  \begin{eqnarray*}
  a_{2+} |n_2) = \left[ n_2+1 \right]_q |n_2+1), \quad 
  a_{2+} |k-1) = 0
  \end{eqnarray*}
  \begin{eqnarray*}
  a_{2-} |n_2) = |n_2 - 1), \quad 
  a_{2-} |0) = 0
  \end{eqnarray*}
  \begin{eqnarray*}
  N_1 |n_1) = n_1 |n_1), \quad 
  N_2 |n_2) = n_2 |n_2)
  \end{eqnarray*}
on a Fock space 
${\cal F} = \{ | n_1 n_2 ) = |n_1) \otimes |n_2) : 
                 n_1,n_2 = 0, 1, \cdots, k-1 \}$ 
of finite dimension (dim~${\cal F} = k^2$). We use here the notation 
  \begin{eqnarray*}
  \left[ x \right]_q = \frac{1-q^x}{1-q} \quad {\hbox{for}} \quad x \in {\bf R}
  \end{eqnarray*}
so that $[n]_q = 1 + q + \cdots + q^{n-1}$ for $n \in N^*$.

We now define the two following linear operators
  \begin{eqnarray*}
  H = {\sqrt {N_1 \left( N_2 + 1 \right) }}
  \end{eqnarray*}
and
  \begin{eqnarray*}
  U_r = \left[ a_{1+} + 
  {\rm exp} \left( {{\rm i}{{\phi}_r \over 2}} \right) {(a_{1-})^{k-1} \over 
  \left[ k-1 \right]_q!} \right]
        \left[ a_{2-} + 
  {\rm exp} \left( {{\rm i}{{\phi}_r \over 2}} \right) {(a_{2+})^{k-1} \over 
  \left[ k-1 \right]_q!} \right]
  \end{eqnarray*}
where the real parameter $\phi_r$ is taken in the form 
  \begin{eqnarray*}
  \phi_r = \pi (k-1) r \quad {\rm with} \quad r \in {\bf R}
  \end{eqnarray*}
and the $q$-deformed factorial is defined by 
  \begin{eqnarray*}
  \left[ n \right]_q! = 
  \left[ 1 \right]_q 
  \left[ 2 \right]_q \cdots 
  \left[ n \right]_q \quad {\rm for} \quad n \in {\bf N}^* 
  \quad {\rm and} \quad \left[ 0 \right]_q! = 1  
  \end{eqnarray*}
The action of $U_r$ on $\cal F$ is easily found to satisfy 
  \begin{equation}
  U_r |n_1 n_2) = |n_1+1, n_2-1) \quad {\hbox{for}} \quad n_1 \not = k-1 \quad 
  {\hbox{and}} \quad n_2 \not = 0
  \end{equation}
and
  \begin{equation}
  U_r |k-1, 0) = {\rm exp} \left( {{\rm i}{{\phi}_r }} \right) 
  |0, k-1)
  \end{equation}
while for $H$ we have
  \begin{equation}
  H |n_1 n_2) = {\sqrt{ n_1 (n_2 + 1) } |n_1 n_2)}
  \end{equation}
By using the Schwinger trick
  \begin{eqnarray*}
  j = {1 \over 2} \left( n_1+n_2 \right), \quad 
  m = {1 \over 2} \left( n_1-n_2 \right)  \quad \Rightarrow \quad 
  |n_1 n_2) = |j + m, j-m) \equiv |j m \rangle
  \end{eqnarray*}
we can rewrite eqs.(4) and (5) as 
  \begin{eqnarray*}
  U_r |jm \rangle = \left[ 1 - \delta (m,j) \right] |j, m+1 \rangle + 
  \delta(m,j)  
  {\rm exp} \left( {{\rm i}{{\phi}_r }} \right)
  |j, -j \rangle 
  \end{eqnarray*}
Similarly, eq.(6) can be rewritten 
  \begin{eqnarray*}
  H |j m \rangle = {\sqrt{ (j+m)(j-m+1) }} |j m \rangle
  \end{eqnarray*}
Furthermore, we have 
  \begin{eqnarray*}
  U_r^{\dagger} | jm \rangle = \left[ 1-\delta(m,-j) \right] | j,m-1 \rangle +
  \delta(m,-j) 
  {\rm exp} \left( - {{\rm i}{{\phi}_r }} \right) 
  | j j \rangle
  \end{eqnarray*}
where ${U_r}^{\dagger}$ stands for the adjoint of $U_r$. For a fixed value of
$k$, we take 
  \begin{eqnarray*}
  2j = k-1 \quad {\rm with} \quad k \in {\bf N} \setminus \{ 0,1 \} 
  \end{eqnarray*}
We can thus have $j = {1 \over 2}, \> 1, \> {3 \over 2}, \> \cdots$. The case 
$j = 0$ corresponds to the limiting situation where $k \to \infty$. 

It is obvious that the operator $H$ is Hermitean and the operator $U_r$ is 
unitary. The action of $U_r$ on ${\cal F}$ is cyclic. As a further property of
$U_r$, we have 
  \begin{eqnarray*}
  \left( U_r \right)^{2j+1} = {\rm exp}( {\rm i} \phi_r ) 
  \end{eqnarray*}  
that reflects the cyclical character of $U_r$. 

Let us introduce the three operators
  \begin{equation}
  J_+ = H           U_r, \quad 
  J_- = U_r^{\dagger} H
  \end{equation}  
and
  \begin{equation}
  J_3 = {1 \over 2} \left( N_1-N_2 \right)
  \end{equation}
It is immediate to check that the action on the state $ | jm \rangle $ 
of the operators defined by eqs.(7) and (8) is given by 
  \begin{eqnarray*}
  J_\pm |jm \rangle = {\sqrt{ (j \mp m)(j \pm m+1) }} |j, m \pm 1 \rangle 
  \end{eqnarray*}
and
  \begin{eqnarray*}
  J_3   |jm \rangle = m |jm \rangle
  \end{eqnarray*}
Consequently, we have the commutation relations 
  \begin{eqnarray*}
  \left[ J_3,J_{\pm} \right] = \pm J_{\pm}, \quad \left[ J_+,J_- \right] = 2J_3 
  \end{eqnarray*}
which correspond to the Lie algebra of the group SU(2). As a result, the
non-deformed Lie algebra su(2) is obtained from two $q$-deformed oscillator 
algebras. 

To close this section,  it is interesting to note that we can generate the
infinite dimensional Lie algebra $W_{\infty}$ from the generators of $A_1$ and 
$A_2$. Indeed, by putting 
  \begin{eqnarray*} 
  U = U_r, \quad V = q^{ N_1 - N_2 }
  \end{eqnarray*}
and
  \begin{eqnarray*}
  T_{ (m_1,m_2) } = q^{m_1m_2} U^{m_1} V^{m_2}
  \end{eqnarray*}
we can prove that 
  \begin{equation}
  \left[ T_m,T_n \right] = -2 \> {\rm i} 
  \sin \left( {2 \pi \over k} m \times n \right) 
  T_{ m+n }
  \end{equation}
where we use the abbreviations 
  \begin{eqnarray*}
  m = \left( m_1,m_2 \right), \quad 
  n = \left( n_1,n_2 \right)
  \end{eqnarray*}
and
  \begin{eqnarray*}
  m+n = \left( m_1+n_1,m_2+n_2 \right), \quad
  m \times n = m_1n_2 - m_2n_1
  \end{eqnarray*}
Equation (9) shows that the operators $T_{\ell}$ span the 
algebra  $W_{\infty}$  introduced  by Fairlie, Fletcher and Zachos [25]. 
This result parallels a similar result obtained in ref.[26] in the study 
of $k$-fermions and of the Dirac quantum phase operator. 

\section{A New Basis for SU(2)}
At this stage, it is important to establish a link with the work by
L\'evy-Leblond [28]. The decomposition (7), in terms of $H$ and $U_r$, 
coincides with the polar
decomposition, described in ref.[28], of the shift operators $J_+$ and $J_-$ of
the Lie algebra su(2). This is easily seen by taking the matrix elements of
$U_r$ and $H$ and by comparing these elements to the ones of the operators 
$\Upsilon$ and $J_T$ in [28]. This yields $H \equiv J_T$~; 
furthermore, by identifying the arbitrary phase $\varphi$ of [28] 
to $\phi_r = 2 \pi j r = \pi (k-1)r$, we obtain that $U_r$ turns
out to be identical to the operator $\Upsilon$ of [28]. 
Equation (7) constitutes an important original result of ref.[28]. 

It is easy to prove that the Casimir operator 
  \begin{eqnarray*}
  J^2 = {1 \over 2} \left( J_+J_- +  
                           J_-J_+ \right) + J_3^2 = H^2 + J_3^2 - J_3
  \end{eqnarray*}
commutes with $U_r$ for any value of $r$. (Note that the commutator 
$[U_r, U_s]$ is different from zero for $r \ne s$.)  Therefore, for fixed $r$,
the commuting set $\{ J^2, U_r\}$ provides us with an alternative to the
familiar commuting set $\{ J^2, J_3 \}$ of angular momentum theory. 
The (complete) set of commuting operators $\{ J^2, U_r\}$ can be easily
diagonalized. This leads to the following result.

{\bf Result}~: The spectra of the operators $U_r$ and $J^2$ are given by
  \begin{equation}
  U_r | j \alpha ; r \rangle = q^{-\alpha} 
      | j \alpha ; r \rangle, \quad 
  J^2 | j \alpha ; r \rangle = j(j+1) 
      | j \alpha ; r \rangle 
  \end{equation}
where 
  \begin{equation}
  |j \alpha ; r \rangle = {1 \over {\sqrt{2j + 1}}} 
  \sum_{m = -j}^j
  q^{\alpha m} 
  |j m \rangle 
  \end{equation}
with the range of values 
  \begin{eqnarray*}
  \alpha = - jr, - jr + 1, \cdots, -jr + 2j, \quad 2j \in {\bf N}
  \end{eqnarray*}
The parameter $q$ in eqs.(10) and (11) is 
  \begin{equation}
  q = \exp \left( {2 \pi {\rm i} \over 2 j + 1 } \right) 
  \end{equation}
(cf.~eq.(2) with $k=2j+1$ for $k \in {\bf N} \setminus \{ 0,1 \}$ and 
$k \to \infty$ for $j=0$). 

It is important to note that in eqs.(10) and (11) the label $\alpha$ goes, 
by step of 1, from $-jr$ to $-jr + 2j$. (It is only for $r=1$ 
that  $\alpha$  goes, by step of 1, from $-j$ to $j$.) The inter-basis 
expansion coefficients 
\begin{eqnarray*}
\langle jm | j \alpha ; r \rangle = {1 \over \sqrt{2 j + 1}} q^{\alpha m} 
\end{eqnarray*}
(with 
$m      = -j , -j  + 1, \cdots,        j$ 
and 
$\alpha = -jr, -jr + 1, \cdots, -jr + 2j$) in eq.(11) define a unitary
transformation that allows to pass from the well-known 
orthonormal standard basis 
$\{ |j m \rangle : 2j \in {\bf N}, \ m = - j, - j + 1, \cdots, j \}$
of the space $\cal F$ to the orthonormal non-standard basis
$B_r = \{ |j \alpha ; r \rangle : 2j \in {\bf N}, \ 
\alpha = - jr, - jr + 1, \cdots, -jr + 2j \}$. 
Then, the expansion  
  \begin{eqnarray*}
  |j m \rangle = {1 \over {\sqrt{2j + 1}}} 
  \sum_{\alpha = -jr}^{-jr + 2j}
  q^{- \alpha m} 
  |j \alpha ; r \rangle  
  \end{eqnarray*}
with 
  \begin{eqnarray*}
  m = - j, - j + 1, \cdots, j, \quad 2j \in {\bf N}
  \end{eqnarray*}
is the inverse of eq.(11).
                           
We thus foresee that it is possible to develop the Wigner-Racah algebra (WRa) 
of the group
SU(2) in the $\{ J^2, U_r \}$ scheme. This furnishes an alternative to the
WRa of SU(2) in the SU(2) $\supset$ U(1) basis corresponding 
to the $\{ J^2, J_3 \}$ scheme. 

\section{A New Approach to the 
           Wigner-Racah Algebra of SU(2)}
In this section, we give the basic
ingredients for the WRa of SU(2) in the $\{ J^2, U_r \}$ scheme. 
The Clebsch-Gordan coefficients (CGc's) adapted to the $\{ J^2, U_r \}$ scheme
are defined from the SU(2) $\supset$ U(1) CGc's adapted to the $\{ J^2, J_3 \}$ 
scheme. The adaptation to the $\{ J^2, U_r \}$ scheme afforded by eq.(11) is
transferred to SU(2) irreducible tensor operators. This yields the
Wigner-Eckart theorem in the $\{ J^2, U_r \}$ scheme.  

\subsection{Coupling and Recoupling Coefficients in the 
             $\{ J^2, U_r \}$ Scheme}
The CGc's or coupling coefficients 
$( j_1 j_2 \alpha_1 \alpha_2 | j \alpha ; r )$ in the $\{ J^2, U_r \}$ scheme 
are simple linear combinations of the SU(2) $\supset$ U(1) CGc's. In fact, we
have
  \begin{eqnarray*}
  \left( j_1 j_2 \alpha_1 \alpha_2 |j \alpha ; r \right) = 
  {1 \over \sqrt{(2j_1 + 1) (2j_2 + 1) (2j   + 1)}} 
  \sum_{m_1=-j_1}^{j_1} 
  \sum_{m_2=-j_2}^{j_2} 
  \sum_{m  =-j  }^{j  } 
  \end{eqnarray*}
  \begin{eqnarray*}
  \times q  ^{  \alpha   m  } 
         q_1^{- \alpha_1 m_1}
         q_2^{- \alpha_2 m_2} \> 
  ( j_1 j_2 m_1 m_2 | j m ) 
  \end{eqnarray*}
where $q$, $q_1$ and $q_2$ are given by eq.(12) in terms of 
      $j$, $j_1$ and $j_2$, respectively. 
The symmetry properties of the coupling coefficients 
$( j_1 j_2 \alpha_1 \alpha_2 | j \alpha ; r )$
cannot be expressed in a simple way
(except the symmetry under the interchange 
$j_1 \alpha_1 \leftrightarrow 
 j_2 \alpha_2$). 
Let us introduce the $f_r$ symbol via 
  \begin{equation}
  f_r\pmatrix{
  j_1     &j_2     &j_3     \cr
  \alpha_1&\alpha_2&\alpha_3\cr
  }
  = (-1)^{2j_3} {1 \over {\sqrt{2j_1+1}}} 
  \left( j_2 j_3 \alpha_2 \alpha_3 | j_1 \alpha_1 ; r \right)^*
  \end{equation}
where the star indicates the complex conjugation. 
Its value is multiplied by the factor 
$(-1)^{j_1 + j_2 + j_3}$ when its two last columns
are interchanged. However, the interchange of two other columns cannot be
described by a simple symmetry property. Nevertheless, the $f_r$ symbol is of
central importance for the Wigner-Eckart theorem in the $\{ J^2 , U_r \}$ 
scheme (see eq.(17) below).  

Following ref.[29], we define a more
symmetrical symbol, namely the ${\bar f}_r$ symbol, through 
  \begin{eqnarray*}
  \bar f_r \pmatrix{
  j_1     &j_2     &j_3     \cr
  \alpha_1&\alpha_2&\alpha_3\cr
  } = 
  {1 \over {\sqrt{(2j_1 + 1) (2j_2 + 1) (2j_3 + 1)} } }
  \sum_{m_1 = -j_1}^{j_1} 
  \sum_{m_2 = -j_2}^{j_2} 
  \sum_{m_3 = -j_3}^{j_3} 
  \end{eqnarray*}
  \begin{equation}
  \times q_1^{ - \alpha_1 m_1 } 
         q_2^{ - \alpha_2 m_2 } 
         q_3^{ - \alpha_3 m_3 } 
  \pmatrix{
  j_1&j_2&j_3\cr
  m_1&m_2&m_3\cr
  }
  \end{equation}
where the parameters $q_i$ are given by eq.(12) 
with $q \equiv q_i$ and $j \equiv j_i$  for $i = 1,2,3$. 
The $3 - jm$ symbol on the right-hand side of eq.(14) is an ordinary 
Wigner symbol for the group SU(2) in the SU(2)$\supset$U(1) basis.
As a matter of
fact, it is possible to 
pass from the $f_r$ symbol to the ${\bar f}_r$ symbol and vice versa by means
of a metric tensor.
The ${\bar f}_r$ symbol is more symmetrical than the $f_r$ symbol.
The ${\bar f}_r$ symbol exhibits the same symmetry properties under 
permutations of its columns as the $3-jm$ Wigner symbol~: Its value 
is multiplied by $(-1)^{j_1 + j_2 + j_3}$ under an odd permutation 
and does not change
under an even permutation. In addition, the orthogonality properties of the
highly symmetrical 
${\bar f}_r$ symbol easily follow from  the corresponding properties of the 
$3 - jm$ Wigner symbol. Thus, we have 
  \begin{equation}
  \sum_{j_3 \alpha_3} (2j_3 +1)   
  \bar f_r \pmatrix{
  j_1     &j_2     &j_3     \cr
  \alpha_1&\alpha_2&\alpha_3\cr
  }^* 
  \bar f_r \pmatrix{
  j_1      &j_2      &j_3     \cr
  \alpha_1'&\alpha_2'&\alpha_3\cr
  } = \delta (\alpha_1' , \alpha_1)
      \delta (\alpha_2' , \alpha_2)
  \end{equation}  
and
  \begin{equation}
  \sum_{\alpha_1 \alpha_2}
  \bar f_r \pmatrix{
  j_1     &j_2     &j_3     \cr
  \alpha_1&\alpha_2&\alpha_3\cr
  } 
  \bar f_r \pmatrix{
  j_1     &j_2     &j_3'     \cr
  \alpha_1&\alpha_2&\alpha_3'\cr
  }^* = {1 \over 2 j_3 + 1} \Delta ( 0 | j_1 \otimes j_2 \otimes j_3 ) 
  \delta (j_3' , j_3) \delta (\alpha_3' , \alpha_3)
  \end{equation}  
where $\Delta ( 0 | j_1 \otimes j_2 \otimes j_3 ) =1$ or 0 according to as the
Kronecker product $(j_1) \otimes (j_2) \otimes (j_3)$ 
contains or does not contain 
the identity irreducible representation (0) of SU(2). 
Observe that the real number $r$ is the same for all the ${\bar f}_r$ symbols 
occurring in eqs.(15) and (16). 

The values of the SU(2) CGc's in the $\{ J^2, U_r \}$ scheme as well
as of the $f_r$ and ${\bar f}_r$ coefficients are not necessarily real 
numbers. For instance, we have the following property under complex conjugation
  \begin{eqnarray*}
  \bar f_r \pmatrix{
  j_1     &j_2     &j_3     \cr
  \alpha_1&\alpha_2&\alpha_3\cr
  }^* =  (-1)^{j_1 + j_2 + j_3} 
  \bar f_r \pmatrix{
  j_1     &j_2     &j_3     \cr
  \alpha_1&\alpha_2&\alpha_3\cr
  }
  \end{eqnarray*}
Hence, the value of the ${\bar f}_r$ coefficient 
is real if $j_1 + j_2 + j_3$ is 
even and pure imaginary if  $j_1 + j_2 + j_3$ is odd. Then, the behavior of the 
${\bar f}_r$ symbol under complex conjugation is completely different as the
one of the ordinary $3-jm$ Wigner symbol. 

Finally, it is worth to mention that the recoupling coefficients 
of the group SU(2) can be
expressed in terms of coupling coefficients of SU(2) in the  
$\{ J^2 , U_r \}$ scheme. For example, the $9-j$ symbol can be expressed 
in terms ${\bar f}_r$ symbols by replacing, in its decomposition in terms of
$3-jm$ symbols, the $3-jm$ symbols by ${\bar f}_r$ symbols. On the other hand,
the decomposition of  the $6-j$ symbol   in terms of  
${\bar f}_r$ symbols requires the introduction of six
metric tensors corresponding to the six arguments of the $6 - j$ symbol. 
These matters may be developed by following the approach initiated in 
refs.[29-32].

\subsection{Wigner-Eckart Theorem in the 
$\{ J^2, U_r \}$ Scheme}
From the spherical components $T^{(k)}_q$ (with $q = -k, -k+1, \cdots, k$) 
of an 
SU(2) irreducible tensor operator ${\bf T}^{(k)}$, we define the 
$2 k + 1$ components 
  \begin{eqnarray*}
  T^{(k)}_\alpha (r) = {1 \over {\sqrt{2k+1}}} \sum_{m=-k}^k q^{\alpha m} 
  T^{(k)}_m 
  \end{eqnarray*}
with
  \begin{eqnarray*}
\alpha = -kr, -kr + 1, \cdots, -kr + 2k, \quad 2k \in {\bf N}
  \end{eqnarray*}
In the $\{ J^2, U_r \}$ scheme, the Wigner-Eckart theorem reads
 \begin{equation}
 \langle \tau_1 j_1 \alpha_1 ; r | T^{(k)}_{\alpha}(r) | 
         \tau_2 j_2 \alpha_2 ; r \rangle = \left( 
         \tau_1 j_1             || T^{(k)}               || 
         \tau_2 j_2 \right) 
 f_r\pmatrix{
 j_1     &j_2     &k     \cr
 \alpha_1&\alpha_2&\alpha\cr
 }
 \end{equation}
where $\left( \tau_1 j_1 || T^{(k)} || \tau_2 j_2 \right)$ 
denotes an ordinary reduced matrix
element. Such an element is basis-independent. Therefore, it does not depend 
on the labels $\alpha_1$, $\alpha_2$ and $\alpha$. On the contrary, 
the $f_r$ coefficient in eq.(17), defined by eq.(13),  
depends on the labels $\alpha_1$, $\alpha_2$ and $\alpha$.   

\section{Concluding Remarks}
In this paper, we have developed a quon approach to the Lie algebra of the
classical (not quantum!) group SU(2). Such an approach leads to the polar
decomposition of the generators $J_+$ and 
                                $J_-$ of SU(2), a decomposition
originally introduced by L\'evy-Leblond [28]. 

The familiar $\{ J^2, J_3 \}$ quantization 
scheme with the (usual) standard spherical basis 
$\{ |jm \rangle : 2j \in {\bf N}, \ m = -j, -j+1, \cdots, j\}$, 
corresponding to
the canonical chain of groups SU(2)$\supset$U(1), is thus replaced by the 
$\{ J^2, U_r \}$ quantization scheme with a (new) basis, 
namely, the non-standard basis    
$B_r 
= \{ |j \alpha ; r \rangle : 2j \in {\bf N}, 
\ \alpha = -jr, -jr+1, \cdots, -jr + 2j\}$. We have given the premises of the
construction of the Wigner-Racah algebra of the group SU(2) in the 
$B_r$ basis. Of course, there exists an infinity of $B_r$ bases due to the fact
that $r \in {\bf R}$. The case $r=1$ probably deserves a special attention. 
We shall give elsewhere a complete development of the Wigner-Racah algebra of
SU(2) in the $B_1$ basis. In particular, the calculation and the
properties, including Regge symmetry properties, of the coupling coefficients 
(${\bar f}_1$ and $f_1$ symbols and CGc's in the $\{ J^2 , U_1 \}$ scheme) 
shall be the object of a forthcoming paper. 

As a further interesting step, it would be interesting to find realizations of
the $B_r$ basis (i) on the sphere $S^2$ for $j$ integer and (ii) on the
Fock-Bargmann spaces (of entire analytical functions) in 1 and 2 dimensions for 
$j$ integer or half of an odd integer. In this respect, the problem of finding
a differential realization of the operator $U_r$ on $S^2$ and of expressing its
eigenfunctions
  \begin{equation}
  \left[ y_r \right]_{\ell \alpha} (\theta, \varphi) 
  = { 1 \over \sqrt{ 2 \ell + 1 } } \sum_{m = -\ell}^{\ell} q^{\alpha m}  
  Y_{\ell m} (\theta, \varphi) 
  \end{equation}
with
  \begin{eqnarray*}
  \alpha = - \ell r, - \ell r + 1, \cdots, - \ell r + 2 \ell, \quad 
   \ell \in {\bf N} 
  \end{eqnarray*}
as special functions is very appealing. (In eq.(18), $Y_{\ell m}$ denotes a
spherical harmonic.) 

\section{Acknowledgments}
One of the authors (M.K.) wishes to thank Jean-Marc L\'evy-Leblond 
(from the Universit\'e Sophia Antipolis, Nice, France) 
for attracting his attention on ref.[28] sixteen years ago.  He also wishes to
thank Jacob Katriel 
(from the Israel Institute of Technology, Haifa, Israel) for
discussions which contributed to clarify the question of the spectrum of the
operator $U_{r=0}$ eight years ago. 
Finally, he would like to thank the organizing committee of
the International Workshop {\em Quantum Groups, Deformations and Contractions} 
for inviting him to give a talk from which the present paper is a by-product.

\end{document}